\documentclass[10pt,doublecolumn,twosided,standalone]{IEEEtran}

\usepackage{blindtext, graphicx}
\usepackage{cite}
\usepackage{verbatim}
\usepackage{fancyhdr} 

\fancyheadoffset{0cm}

\fancypagestyle{plain}{%
	\fancyhf{}%
	\fancyhead[R]{\thepage}%
}

\ifCLASSINFOpdf
\else
\fi

\usepackage{amsmath}

\usepackage{blindtext}
\usepackage{multicol}
\usepackage{graphicx}
\usepackage{caption}

\usepackage[latin1]{inputenc}
\usepackage{amsfonts}
\usepackage{amssymb}
\usepackage{graphicx}
\usepackage{epspdfconversion}

\newcommand{\RNum}[1]{\uppercase\expandafter{\romannumeral #1\relax}}

\newcommand{\diag}{{\mathrm{diag}}}
\usepackage{algorithmicx}
\usepackage[ruled]{algorithm2e}

\usepackage{12many}
\title{Analysis of One-Bit Quantized Precoding for the Multiuser Massive MIMO Downlink}

\author{Amodh Kant Saxena$^{1}$, Inbar Fijalkow$^2$, A. Lee Swindlehurst$^1$
	\\
	$^1$ Center for Pervasive Communications and Computing, University of California Irvine, Irvine, CA 92697, USA, \{aksaxena, swindle\}@uci.edu\\
	$^2$ ETIS, UMR 8051 / ENSEA, Universit\'e Cergy-Pontoise, CNRS, F-95000 Cergy, France, inbar.fijalkow@ensea.fr}

\begin{document}

\maketitle
\thispagestyle{empty}
\textbf{Abstract -
We present a mathematical analysis of linear precoders for downlink massive MIMO multiuser systems that employ one-bit digital-to-analog converters at the basestation in order to reduce complexity and mitigate power usage. The analysis is based on the Bussgang theorem, and applies generally to any linear precoding scheme. We examine in detail the special case of the quantized zero-forcing (ZF) precoder, and derive a simple asymptotic expression for the resulting symbol error rate at each terminal. Our analysis illustrates that the performance of the quantized ZF precoder depends primarily on the ratio of the number of antennas to the number of users, and our simulations show that it can outperform the much more complicated maximum likelihood encoder for low-to-moderate signal to noise ratios, where massive MIMO systems are presumed to operate. We also use the Bussgang theorem to derive a new linear precoder optimized for the case of one-bit quantization, and illustrate its improved performance.}


\section{Introduction}
Massive MIMO involves the use of many, perhaps hundreds, of antennas at the base station (BS) of a wireless network, and can potentially provide large increases in capacity via spatial multiplexing \cite{Larsson2014MassMIMO}. In a multi-user (MU) scenario, the massive MIMO BS typically serves a number of users much smaller than the number of antennas, and hence a large number of degrees-of-freedom can be offered to each user.  This can in turn lead to improved robustness and correspondingly high data rates \cite{Larsson2014MassMIMO,Fusek2013ScalingMIMO,LS2014MassiveMIMO}.
	
Under favorable propagation conditions, the user channels become asymptotically orthogonal as the number of antennas grows, and simple linear precoding at the BS can be used to invert the channel without noise enhancement. Many studies consider Maximal Ratio Transmission (MRT) \cite{TKLo1999MaximumRatio} or Zero Forcing (ZF) precoders \cite{Shamai2008ZFPrecoding}, and asymptotic results from random matrix theory show how an increasing number of antennas can result in a dramatic increase in downlink capacity \cite{Marzetta2013ZFBeamforming} even for these simple precoding schemes.
	
While the benefits of massive MIMO at the BS grow with the number of antennas, so do the resulting power consumption and hardware costs. While one can scale down the transmit power with an increase in the number of antennas in order to maintain a certain level of performance (e.g., due to beamforming gain), there are certain sources of fixed power consumption at the circuit level that cannot be reduced and these sources will lead to an increase in power as the number of antennas is increased \cite{Kang2013CircuitPower}. More important than this is the issue of energy efficiency; a standard RF implementation requires highly linear amplifiers that must as a result be operated with considerable power back-off, which severely limits the overall energy efficiency of the system. The more RF chains, the less and less efficient the system is.

One approach to addressing this problem in the massive MIMO downlink is the use of hybrid analog and digital RF front ends, which employ fewer RF chains in favor of an analog beamforming (precoding) network that is deployed after the digital-to-analog converters (DACs) \cite{Han2015Hybrid5G, Roh2014MmWaveBeamforming, Heath2013HybridPrecode}. However, this approach does not scale well for wideband systems, since a different phase-shift network is needed for each frequency band. Instead, we focus on another approach that has gained attention recently, namely the use of low-resolution DACs for each antenna and RF chain; in particular, we will investigate the simplest possible case involving one-bit DACs. Using one-bit ADCs/DACs considerably reduces power consumption, which grows linearly with increases in bandwidth and sampling rate, and exponentially in the number of quantization bits \cite{walden1999analog,Svensson2006ADCpower,singh2009multi}. More importantly for the downlink, it eliminates the need for highly linear amplifiers and back-off operation, which further reduces circuit complexity and dramatically improves energy efficiency. As we will show in this paper, the severe distortion caused by the one-bit DACs can be mitigated by proper signal processing, and the impact is not too significant in the low- to mid-SNR ranges where massive MIMO systems will likely operate.

Most of the work on one-bit quantization for wireless communication systems has focused on the uplink, where the BS employs one-bit analog-to-digital converters (ADCs). Single-antenna studies of the impact of one-bit ADCs can be found in \cite{Amine2008onebitRayleigh, Amine2010onebitChannelEstimation, Amine2013BOConebit}. More recently, their use in MIMO systems has been considered, and the resulting work has focused primarily on channel estimation and information theoretic rate analyses \cite{Zhang2015MixedADC, jacobsson2015one, Yongzhi2016ChannelAcheivableRateOneBit, jianhua2015capacity,Nossek2007UWB1bit,Nossek2012CapacityMIMOQuant}.While there has been considerable research on downlink precoding for massive MIMO (see e.g.,\cite{Zhao2014ZFPrecode}-\cite{Debbah2013Cellular}) very little has been reported on the impact of low-resolution DACs on transmit processing. In \cite{Nossek2009Transmit}, transmit optimization for the case of flat fading MIMO systems with low resolution DACs is addressed. The mean squared error between the received symbol and the symbol vector input to the transmitter is minimized to find optimum quantizer levels, transmit matrix and scalar receiver. In \cite{Jedda2016MinimumBER}, a precoding technique is introduced which aims to minimize the inter-user interference and quantization noise introduced by using a look-up table for all possible transmit sequence combinations. This paper also introduced a novel minimum Bit Error Rate (BER) performance metric. In \cite{Usman2016MMSEprecoder}, a two stage precoder is proposed, which comprises a digital precoder, and an analog precoder implemented after the quantizer. The precoders are optimized by minimizing the mean square error between the transmit vector and the receive vector. An iterative algorithm is utilized in the optimization problem.

In this paper, we study the impact of one-bit DACs on linear precoding for the massive MIMO downlink. We presented a preliminary analysis of this problem in \cite{LS2016OneBitZF} using a different approach. To focus on the performance degradation due to quantization, we assume that the BS has perfect channel state information, although this additional error source would have to be accounted for in a full analysis. Using the Bussgang theorem \cite{Bussgang1952GaussianCorrelation} to model the second-order statistics of the quantization noise introduced by the DACs, we provide a closed-form expression for the signal to quantization, interference and noise ratio (SQINR), which we use to deduce the symbol error rate for each terminal in the network. We then focus on the special case of the zero-forcing (ZF) precoder and use asymptotic arguments to obtain an even simpler expression. Our analysis illustrates that the performance of the quantized ZF precoder depends primarily on the ratio of the number of antennas to the number of user terminals, and our simulations show that it can outperform the much more complicated maximum likelihood encoder for low-to-moderate signal to noise ratios, where massive MIMO systems are presumed to operate. Finally, using our insights from the asymptotic analysis of the ZF precoder, we design a modified precoder that attempts to achieve the benefits of the ZF precoder even in the non-asymptotic case.

The paper is organized as follows. Section \RNum{2} introduces the one-bit downlink model, and describes both direct non-linear maximum likelihood precoding and the simpler quantized linear precoding approach. The SQINR  performance of a general one-bit quantized linear precoder is then analyzed in Section \RNum{3}, and the approximate Symbol Error Rate (SER) for each user is derived. Section \RNum{4} focuses on the special case of ZF precoding in the asymptotic regime where the number of BS antennas $M$ and user terminals $K$ become large, leading to a simpler and more insightful expression. In Section \RNum{5}, we introduce the Bussgang adapted precoding algorithm, which attempts to remove the interuser interference for non-asymptotic values of $M$ and $K$. Simulation results comparing the various algorithms are presented throughout the paper.

	
\section{One-Bit Downlink System Model}

\subsection{Mathematical Notation and Assumptions}

In what follows, uppercase boldface letters, $\mathbf{A}$, indicate a matrix, with $[\mathbf{A}]_{kl}$ and $a_{kl}$ interchangeably denoting the element at the $k^{th}$ row and $l^{th}$ column. Lower boldface letters, $\mathbf{a}$, indicate a column vector, with $a_k$ denoting the $k^{th}$ element of the column vector. The symbols $(.)^*$, $(.)^T$ and $(.)^H$ denote the complex conjugate, matrix transpose and the transpose-conjugate of the argument respectively. We will use $\text{diag}(\mathbf{C})$ to denote the square matrix whose main-diagonal elements are equal to those of the square matrix $\mathbf{C}$, and whose other entries are all zero. With a vector argument, Diag($\mathbf{c}$) denotes the diagonal matrix whose main diagonal is composed of the elements of vector $\mathbf{c}$.

We assume a flat-fading downlink scenario in which an $M$-antenna BS is attempting to send QPSK symbols $s_k$ to $k=1,\cdots,K$ single-antenna users over the $K \times M$ channel $\mathbf{H}$. The BS transmits an $M\times 1$ vector $\sqrt{\rho}\mathbf{x}$, where $\sqrt{\rho}$ is a fixed gain and the elements of $\mathbf{x}$ are constrained to be equal to $\pm 1 \pm j$ due to the use of one-bit quantization of the in-phase and quadrature components of the signal at the BS. Let $r_k$ be the signal received by user $k$, and define $\mathbf{r}=[r_1 \; \cdots \; r_K]^T$ so that we can write the overall system model as
\begin{equation}\label{rhxn}
\mathbf{r} = \sqrt{\rho}\mathbf{Hx} + \mathbf{n} \; ,
\end{equation}
where the $K \times 1$ vector $\mathbf{n}$ represents a vector of independent Gaussian noise terms of variance $\sigma_n^2$ at each user. For the downlink, the BS designs the vector $\mathbf{x}$ such that the elements of $\mathbf{r}$ can be correctly decoded as the appropriate QPSK symbols in the vector $\mathbf{s}=[s_1 \; \cdots \; s_K]^T$. The QPSK symbols for different users are assumed to be zero-mean and independent with power $\sigma_s^2$: $E(\mathbf{s}\mathbf{s}^H)=\sigma^2_s\mathbf{I}_K$. The assumption that the BS transmits QPSK symbols can be interpreted to mean that the individual users also employ one-bit ADCs, although this is not strictly necessary.

The $k^{th}$ row of channel matrix $\mathbf{H}$ is denoted as $\mathbf{h}_k$, and represents the channel to user $k$. For our analysis, we will assume that the channel matrix is given by
\begin{equation}
\mathbf{H}=\text{Diag}(\sigma_1, \sigma_2, \ldots, \sigma_K)\mathbf{\tilde{H}} \; ,
\end{equation}
where the elements of $\mathbf{\tilde{H}}$ are complex Gaussian random variables, whose real and imaginary parts are both iid Gaussian random variables with zero-mean and unit-variance, and the parameters $\sigma_1, \sigma_2, \ldots, \sigma_K$ represent potentially different path losses for each user. As mentioned above, we will assume that the channel $\mathbf{H}$ is known at the BS in order to isolate the impact of the quantization and noise.

\subsection{ML Precoding}

Ideally, in the absence of noise, one might attempt to design a general non-linear precoder providing $\mathbf{x}$ such that $\mathbf{H}\mathbf{x} = \mathbf{s}$.  However, due to the finite alphabet constraint imposed by the one-bit DACs, one would have to find such an $\mathbf{x}$ with QPSK entries, which in general will prevent achieving $\mathbf{H}\mathbf{x} = \mathbf{s}$ with equality. Instead, assuming Gaussian noise, one might choose to implement a maximum likelihood (ML) encoder, which attempts to solve \cite{peelVectorPerturb2005}
\begin{equation}		
\mathbf{x}=\text{arg }\min_{\mathbf{v} \in \mathcal{S}^M} ||\mathbf{s}-\mathbf{H}\mathbf{v}||^2 \; ,
\label{eq_ml}
\end{equation}
where $\mathcal{S}=\{1+j,1-j,-1+j,-1-j\}$ is the set of QPSK constellation points. However, in general, (\ref{eq_ml}) requires on exhaustive search of order $O(4^M)$, which is prohibitively expensive even for relatively small values of $M$, let alone in the massive MIMO case. Even a sphere-based encoder \cite{Pohst1985Lattice} would be too complex for large values of $M$, and in this case would require extra care since the matrix $\mathbf{H}$ has many more columns than rows. In such cases, one should transform~(\ref{eq_ml}) to
\begin{equation}		
\mathbf{x}=\arg \min_{\mathbf{v} \in \mathcal{S}^M} ||\mathbf{D}(\mathbf{z}-\mathbf{v})||^2 \; ,
\label{eq_ml2}
\end{equation}
where $\mathbf{D}$ is the upper triangular matrix obtained by the Cholesky factorization of $\mathbf{G}=\mathbf{H}^H\mathbf{H}+\alpha\mathbf{I}_M$, $\mathbf{z} = \mathbf{G}^{-1}\mathbf{H}^H\mathbf{s}$ and $\alpha$ is a small regularization parameter as explained in \cite{Cui2005Sphere}. Though less complex than direct ML encoding, the generalized sphere encoder still has a complexity exponential in $M-K$, which is again costly in the massive MIMO case.

We note here that, for the case where the elements of the desired vector $\mathbf{s}$ are themselves drawn from a finite alphabet (QPSK here), the ML encoder over-constrains the problem by attempting to force $\mathbf{Hx}$ to be close to $\mathbf{s}$, when in fact all that is necessary is that its elements lie within the correct decision regions so that the users can properly decode them as the desired constellation points $s_k$. The noise-free data $\mathbf{Hx}$ can in principle be far away from $\mathbf{s}$ and still be decoded correctly; in fact, often the farther $\mathbf{Hx}$ is away from $\mathbf{s}$, the farther the received signal is from the decision boundaries and hence the more resilient to noise. So we might expect that ML encoding may not give optimal performance in this case, and in fact we will demonstrate this fact later in the paper.

\subsection{One-bit Quantized Linear Precoding}
\label{OneBitPrecodeModel}
\begin{figure}
\centering
\includegraphics[width=0.5\textwidth]{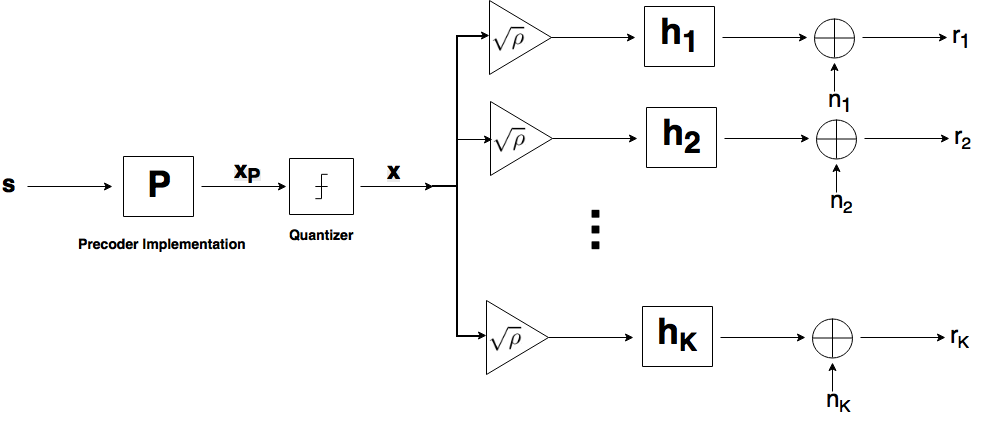}
\caption{System Model}
\label{fig_model}
\end{figure}

As an alternative to ML encoding, we will study the performance of a very simple approach in which the output of a standard linear precoder is quantized by one-bit DACs prior to transmission. In particular, assuming a linear precoding matrix $\mathbf{P}$, the transmitted signal is $\mathbf{x} = \mathbb{Q}(\mathbf{Ps})$, where the one-bit quantization operation is defined as
\begin{equation}
\mathbb{Q}(\mathbf{a})=\text{sign}(\Re(\mathbf{a}))+j \text{sign}(\Im(\mathbf{a})),
\end{equation}
with $\Re(\mathbf{\cdot})$ representing the real part, $\Im(\mathbf{\cdot})$ the imaginary part, and $\text{sign}(\cdot)$ the sign of their arguments. Figure~\ref{fig_model} gives a graphical view of the assumed system, whose output is thus given by
\begin{equation}\label{rec_vec}
\mathbf{r}= \sqrt{\rho}\mathbf{H}\mathbb{Q}(\mathbf{Ps})+\mathbf{n} \; .
\end{equation}
In what follows, we will assume that $\rho=\frac{\rho_0}{M}$, where, $\rho_0$ is defined to be the transmit SNR.

\section{Bussgang Analysis of One-Bit Quantized Precoding}

Let $\mathbf{x}_P=\mathbf{Ps}$ represent the precoded vector before quantization.
In this section, we use the Bussgang decomposition to analyze the impact of the quantization on
the signal of interest and to quantify the level of quantization noise.
This will allow us in the sequel to approximate the SQINR.

\subsection{Bussgang Decomposition}

The one-bit quantization operation on the precoded vector $\mathbf{x}_P$ is modeled here using the Bussgang theorem \cite{Bussgang1952GaussianCorrelation}.
We assume that the vector of QPSK symbols $\mathbf{s}$ is random with zero mean and covariance matrix $\sigma^2_s\mathbf{I}_K$. Although
this implies that $\mathbf{x}_P$ is not strictly Gaussian, each element of $\mathbf{x}_P$ is formed as a result of a linear mixture of
the $K$ $i.i.d.$ elements of the vector $\mathbf{s}$, the Gaussian assumption is fulfilled for large enough $K$. We thus apply the Bussgang
theorem to write
\begin{equation}
\mathbf{x}:=\mathbb{Q}(\mathbf{x}_P)=\mathbf{F}\mathbf{x}_P+\mathbf{q} \; ,
\end{equation}
where $\mathbf{F}$ is chosen to satisfy $\mathbf{R_{x_Pq}}=E(\mathbf{x}_P\mathbf{q}^H)=\mathbf{0}$. The Bussgang theorem
provides a linear representation of the quantization that is statistically equivalent up to the second moments of the data.
To define the decomposition, we have
\begin{equation}
\begin{split}
\mathbf{R_{xx_P}}=E(\mathbf{x}\mathbf{x}_P^H)=& E(\{\mathbf{Fx_P}+\mathbf{q}\}\mathbf{x}_P^H)\\
= \mathbf{F}\mathbf{R_{x_Px_P}} \; ,
\end{split}
\label{eq_f}
\end{equation}
where
\begin{equation}\label{eq_Rxpxp}
\mathbf{R_{x_Px_P}}=\sigma^2_s\mathbf{P}\mathbf{P}^H.
\end{equation}
Note that the $M\times M$ matrix in (\ref{eq_Rxpxp}) is rank deficient, and thus
$\mathbf{F}$ cannot be solved for directly as in \cite{Amine2013BOConebit}. We will see shortly that
a unique expression for $\mathbf{F}$ is unnecessary, and that~(\ref{eq_f}) is sufficient.

Under the mutual Gaussian assumption between the components of $\mathbf{x}_P$, the inter-correlation between the one-bit quantized $x_i$ and unquantized $x_{P,j}$ is equal to the normalized inter-correlation of the unquantized signals reduced by a factor of $\sqrt{2/\pi}$, as in [33]:
\[
E(x_i x_{P,j}^*)= \sqrt{\frac{2}{\pi}}
\frac{ E(x_{P,i} x_{P,j}^*)}{\sigma_{x_{P,i}}} \; ,
\]
where $\sigma_{x_{P,i}} = \{E(x_{P,i} x_{P,i}^*)\}^{\frac{1}{2}}$. In matrix form
this yields
\begin{equation}
\begin{split}
\mathbf{R_{xx_P}}=&\sqrt{\frac{2}{\pi}}\left\lbrace \diag(E(\mathbf{x}_P\mathbf{x}_P^H)) \right\rbrace^{-\frac{1}{2}}E(\mathbf{x}_P\mathbf{x}_P^H)\\ =&\sqrt{\frac{2}{\pi}}\sigma_s\left\lbrace \diag(\mathbf{P}\mathbf{P}^H)\right\rbrace^{-\frac{1}{2}}\mathbf{P}\mathbf{P}^H.
\end{split}
\end{equation}
Thus, from~(\ref{eq_f}),
\begin{equation}
\mathbf{F}\mathbf{PP}^H=\frac{1}{\sigma_s}\sqrt{\frac{2}{\pi}}\left\lbrace \diag(\mathbf{P}\mathbf{P}^H)\right\rbrace^{-\frac{1}{2}}\mathbf{PP}^H \; ,
\end{equation}
and since $\mathbf{P}$ is full column rank, we have
\begin{equation}
\mathbf{F}\mathbf{P}=\frac{1}{\sigma_s}\sqrt{\frac{2}{\pi}}\left\lbrace \diag(\mathbf{P}\mathbf{P}^H)\right\rbrace^{-\frac{1}{2}}\mathbf{P}.
\label{eq_FP}
\end{equation}
We will see in Section~\ref{BussgangAnalysis} that this expression for $\mathbf{F}\mathbf{P}$ will be sufficient for our analysis, even if $\mathbf{F}$ can not be fully defined.

\medskip
It is also useful to derive here the covariances of the quantization noise $\mathbf{q}$ and the data vector $\mathbf{x}$ after quantization. Using the arcsin law, for a hard limiting one-bit quantizer, we have \cite{ProbabilityBook1991}
\begin{equation}
E(x_ix_j^*)=\frac{2}{\pi}\text{arcsin}\left(\frac{E(x_{P,i}x_{P,j}^*)}{\sigma_{x_{P,i}}\sigma_{x_{P,j}}}\right),
\end{equation}
which implies
\begin{equation}
\mathbf{R_{xx}}=\frac{2}{\pi}\text{arcsin}\left[\{\text{diag}(\mathbf{R_{x_Px_P}})\}^{-\frac{1}{2}}\mathbf{R_{x_Px_P}}\{\text{diag}(\mathbf{R_{x_Px_P}})\}^{-\frac{1}{2}}\right].
\end{equation}
For the quantization noise vector $\mathbf{q}$, we have
\begin{equation}
\begin{split}
\mathbf{R_{qq}}=& \mathbf{R_{xx}}-\mathbf{F}\mathbf{R_{x_{P}x}}\\&+\mathbf{F}\mathbf{R_{x_{P}x_{P}}}\mathbf{F}^H-\mathbf{R_{xx_{P}}}\mathbf{F}^H
\\=& \mathbf{R_{xx}}-\mathbf{FP}\mathbf{P}^H\mathbf{F}^H\sigma^2_s
\\=&\frac{2}{\pi}[ \arcsin \lbrace \{\diag( \mathbf{PP}^H)\}^{-\frac{1}{2}}\mathbf{PP}^H \{\diag( \mathbf{PP}^H)\}^{-\frac{1}{2}}\rbrace\\&-\{\diag( \mathbf{PP}^H)\}^{-\frac{1}{2}}\mathbf{PP}^H \{\diag( \mathbf{PP}^H)\}^{-\frac{1}{2}} ].
\end{split}
\label{eq_rqq}
\end{equation}

\subsection{Impact on the Signal of Interest}
\label{BussgangAnalysis}

Let $\tilde{\mathbf{s}}$ be the noiseless received signal vector
\begin{equation}
\tilde{\mathbf{s}} = \sqrt{\rho}\mathbf{H}\mathbf{x} =
\frac{\sqrt{\rho_0}}{\sqrt{M}} \mathbf{H}\mathbf{x}.
\end{equation}
The cross-correlation between the received $\mathbf{\tilde{s}}$ and desired $\mathbf{s}$ is
\begin{equation}
\begin{split}
\label{eq_RssHFP}
\mathbf{R_{\tilde{s}s}}=&\frac{\sqrt{\rho_0}}{\sqrt{M}}E(\mathbf{H}\mathbf{x}\mathbf{s}^H)\\=&\frac{\sqrt{\rho_0}}{\sqrt{M}}\mathbf{H}E\lbrace(\mathbf{F}\mathbf{P}\mathbf{s}+\mathbf{q})\mathbf{s}^H\rbrace\\ =& \frac{\sqrt{\rho_0}\sigma^2_s}{\sqrt{M}}\mathbf{HFP} \; ,
\end{split}
\end{equation}
where (\ref{eq_RssHFP}) results because
\begin{equation}
E(\mathbf{x_Pq}^H)=\mathbf{P}E(\mathbf{sq}^H)=\mathbf{0}
\end{equation}
Since $\mathbf{P}$ is full column rank, we can say that,
\begin{equation}
E(\mathbf{sq}^H)=\mathbf{0}
\end{equation}

In the sequel, we denote $\mathbf{G} = \mathbf{HFP}$.
Using (\ref{eq_FP}), $\mathbf{G} = \sqrt{\frac{2}{\pi}}\frac{1}{\sigma_s} \mathbf{H}
\left\lbrace \diag(\mathbf{P}\mathbf{P}^H)\right\rbrace^{-\frac{1}{2}}\mathbf{P}$
and
\begin{equation}
\mathbf{R_{\tilde{s}s}}=\sqrt{\frac{2}{\pi}}\frac{\sqrt{\rho_0}\sigma_s}{\sqrt{M}}\mathbf{H}
\left\lbrace \diag(\mathbf{P}\mathbf{P}^H)\right\rbrace^{-\frac{1}{2}}\mathbf{P}.
\label{eq_rss}
\end{equation}
Equation~(\ref{eq_rss}) shows that, for any full rank precoder, the impact of the
one-bit quantization on the signal of interest is the diagonal matrix
$\left\lbrace \diag(\mathbf{P}\mathbf{P}^H)\right\rbrace^{-\frac{1}{2}}$ and a scalar
factor $\sqrt{\frac{2}{\pi}}$.

\subsection{SQINR and Probability of Error}

Using the Bussgang decomposition, the received vector after quantization can be represented as
\begin{equation}
\begin{split}
\mathbf{r}=&\frac{\sqrt{\rho_0}}{\sqrt{M}}\mathbf{H}(\mathbf{FPs}+\mathbf{q})+\mathbf{n}\\=&\frac{\sqrt{\rho_0}}{\sqrt{M}}\mathbf{Gs}+\frac{\sqrt{\rho_0}}{\sqrt{M}}\mathbf{Hq}+\mathbf{n} \; .
\end{split}
\end{equation}
Letting $\mathbf{d}=\mathbf{Hq}$, we denote the covariance matrix
of the received quantized noise as
\begin{equation}
\mathbf{R_{dd}}=\mathbf{HR_{qq}H}^H.
\label{eq_rdd}
\end{equation}
With these definitions, the SQINR experienced by user $k$ for an arbitrary linear
precoder $\mathbf{P}$ whose output is one-bit quantized can be expressed as
\begin{equation}
\begin{split}
SQINR_k=&\frac{\rho_0\frac{|g_{kk}|^2\sigma^2_s}{M}}{\rho_0\sum_{l=1,l\neq k}^{K} \frac{|g_{kl}|^2\sigma^2_s}{M} +\rho_0\frac{[\mathbf{R_{dd}}]_{kk}}{M}+\sigma^2_n} \; ,
\end{split}
\end{equation}
where $\rho_0\sum_{l=1,l\neq k}^{K} \frac{|g_{kl}|^2\sigma^2_s}{M}$ accounts for multi-user interference and $\rho_0\frac{[\mathbf{R_{dd}}]_{kk}}{M}$ for quantization noise.
With the assumption of equally likely Gray-mapped QPSK signaling, using the nearest neighbour approximation, we can calculate the probability of a decoding error for user $k$ as
\begin{equation}
\begin{split}
P_e=& Pr(\mathbb{Q}(r_k)\neq s_k) \simeq 2Q(\sqrt{SQINR_k}) \\=& 2Q\left(\sqrt{\frac{\rho_0\frac{|g_{kk}|^2\sigma^2_s}{M}}{\rho_0\sum_{l=1,l\neq k}^{K} \frac{|g_{kl}|^2\sigma^2_s}{M} +\rho_0\frac{[\mathbf{R_{dd}}]_{kk}}{M}+\sigma^2_n}}\right).
\end{split}
\end{equation}

\section{Asymptotic Performance of the One-bit Quantized Zero-Forcing Precoder}

The previous section provides a closed-form expression for the SQINR for any
one-bit quantized linear precoder $\mathbf{P}$. To get additional insight into
the impact of the one-bit DACs, here we focus on the special case of the
zero-forcing (ZF) precoder defined by
\begin{equation}
\mathbf{P}=\mathbf{H}^H(\mathbf{HH}^H)^{-1} \; .
\label{eq_ZF}
\end{equation}
In addition, we will further simplify the resulting expressions by adopting a
massive MIMO assumption and letting both $M$ and $K$ be large \cite{LS2014MassiveMIMO}.

\subsection{Approximations for the Asymptotic Case}

In our asymptotic analysis, we let $M$ and $K$ grow large while maintaining a finite value for the ratio $\frac{M}{K}$ that is greater than $1$.
In what follows, we recall and extend some results
on the asymptotic behaviour of the matrix $(\mathbf{HH}^H)^{-1}$ needed for analyzing the ZF precoder.
As mentioned earlier, the channel matrix is assumed to be described as
\begin{equation}
\mathbf{H}=\mathbf{\Sigma}\mathbf{\tilde{H}} \; ,
\end{equation}
where $\mathbf{\Sigma} = \text{Diag}{(\sigma_1,...,\sigma_K)}$ denotes the individual channel gains, and
we assume that the elements of $\mathbf{\tilde{H}}$ are $i.i.d.$ circularly symmetric Gaussian random variables with $\Re(\tilde{h}_{ij})\sim\mathcal{N}(0,1)$ independent of $\Im(\tilde{h}_{ij})\sim\mathcal{N}(0,1),\quad \forall i=1, 2, \ldots, K$ and $j=1,2, \ldots, M$.

It is shown in \cite{goreCL2002} that $\mathbf{Z}= 2\mathbf{{H}{H}}^H$
is a complex Wishart matrix (see \cite{muirhead1982}) with distribution
\begin{equation}
\label{eq_wishart}
\mathbf{W}_K(M,\mathbf{\Sigma}^2; \mathbf{Z})=\frac{|\mathbf{Z}|^\frac{M-K-1}{2}\exp\left({-\frac{1}{2}\text{trace}(\mathbf{\Sigma^{-2}\mathbf{Z}})})\right)}{2^\frac{MK}{2}\Gamma_K(\frac{M}{2})|\mathbf{\Sigma^2}|^\frac{M}{2} }
\end{equation}
where
\[
\Gamma_K(M) = \pi^{\frac{K(K-1)}{4}}\prod_{l=1}^{K}\Gamma(M+\frac{1-l}{2}) \; ,
\]
is the Gamma function. In our case, $\mathbf{\Sigma}$ is diagonal so that
\[
(\det{2(\mathbf{\Sigma})})^{2M} = 2^{2KM}\prod_{i=1}^{K}\sigma^{2M}_i \; .
\]
The variance of the elements of $\mathbf{Y}=\mathbf{Z}^{-1}$, due to the property of the Wishart distribution \cite{muirhead1982}, is given by
\[
Var(y_{ij})=\frac{1}{(2M-K)(2M-K-1)(2M-K-3)\sigma^2_i\sigma^2_j}
\]
for $i\ne k$ and
\[
Var(y_{ii})=\frac{2}{(2M-K-1)^2(2M-K-3)\sigma^4_i}
\]
otherwise. For asymptotic values of $M$, the variance goes to zero for all elements of $\mathbf{Z}^{-1}$, and thus a deterministic value is achieved in the limiting case.

In the asymptotic case, the rows of $\mathbf{H}$ become quasi-orthogonal and
the diagonal terms have been studied in \cite{goreCL2002}
(or \cite{hochwaldAllerton2002} when $\mathbf{\Sigma} = \mathbf{I}_K$).
Given the Wishart distribution (\ref{eq_wishart}), we have
\begin{equation}
(\mathbf{\tilde{H}\tilde{H}}^H)^{-1} \xrightarrow[M\to\infty]{} \frac{1}{2(M-K)}\mathbf{I}_K
\label{eq_HHtilde_inverse}
\end{equation}
and
\begin{equation}
(\mathbf{HH}^H)^{-1}\xrightarrow[M\to\infty]{}\frac{1}{2(M-K)}\mathbf{\Sigma}^{-2}.
\label{eq_HHinverse}
\end{equation}

From (\ref{eq_ZF}) and (\ref{eq_HHinverse}),
\begin{equation}
\mathbf{P} =\mathbf{\tilde{H}}^H \mathbf{\Sigma} \lbrace \mathbf{\Sigma}\mathbf{\tilde{H}}\mathbf{\tilde{H}}^H \mathbf{\Sigma} \rbrace^{-1} \; \;
\xrightarrow[M\to\infty]{} \; \; \frac{1}{2(M-K)}\mathbf{\tilde{H}}^H \mathbf{\Sigma}^{-1} \; .
\label{eq_asymptot_p}
\end{equation}
From (\ref{eq_asymptot_p}),
\begin{equation}
\mathbf{P}\mathbf{P}^H =
\mathbf{H}^H(\mathbf{H}\mathbf{H}^H)^{-2} \mathbf{H}\; \;
\xrightarrow[M\to\infty]{} \; \; \frac{1}{4(M-K)^2}\mathbf{\tilde{H}}^H \mathbf{\Sigma}^{-2}\mathbf{\tilde{H}}.
\end{equation}
In what follows, we use these asymptotic approximations to analyze the one-bit quantized ZF precoder.

\subsection{Asymptotic Received Downlink Signal}
\label{AsymptSigl}

Using the results of the previous section, we have
\begin{equation}
\begin{split}
\mathbf{F}\mathbf{P}=&\frac{1}{\sigma_s}\sqrt{\frac{2}{\pi}}\left\lbrace \diag(\mathbf{P}\mathbf{P}^H)\right\rbrace^{-\frac{1}{2}}\mathbf{P}\\
\xrightarrow[M\to\infty]{} \; \; &\frac{1}{2\sigma_s(M-K)}\sqrt{\frac{2}{\pi}}\left\lbrace \text{diag}(\mathbf{PP}^H)\right\rbrace^{-\frac{1}{2}} \mathbf{\tilde{H}}^H\mathbf{\Sigma}^{-1}
\end{split}
\end{equation}
\begin{equation}
\begin{split}
\mathbf{G}=&\mathbf{HFP}\\\xrightarrow[M\to\infty]{} \; \; &\frac{1}{2\sigma_s(M-K)}\sqrt{\frac{2}{\pi}}\mathbf{\Sigma\tilde{H}}\left\lbrace \text{diag}(\mathbf{PP}^H)\right\rbrace^{-\frac{1}{2}} \mathbf{\tilde{H}}^H\mathbf{\Sigma}^{-1}.
\end{split}
\label{eq_Gasymptotic}
\end{equation}
Since the columns of $\mathbf{H}$ become quasi-orthogonal as $K$ becomes large, the $M\times M$ rank $K$ matrix, $\mathbf{PP}^H=\frac{1}{4(M-K)^2}\mathbf{\tilde{H}}^H \mathbf{\Sigma}^{-2}\mathbf{\tilde{H}}$, asymptotically becomes a diagonal matrix with $K$ non-zero diagonal elements, and these non-zero values can be approximated as:
\begin{equation}
[\mathbf{PP}^H]_{kk} \xrightarrow[K\to\infty]{} \frac{\sum_{i=1}^{K}\frac{1}{\sigma^2_i}}{2(M-K)^2}.
\label{eq_pph_asymptotic}
\end{equation}
Moreover, the non-zero diagonal values correspond to eigenvectors that lie in the size $K$ subspace spanned by $\mathbf{H}$.
Thus, using (\ref{eq_pph_asymptotic}) and (\ref{eq_HHtilde_inverse}) in
(\ref{eq_Gasymptotic}),
\begin{equation}
\begin{split}
\mathbf{\Sigma\tilde{H}}\left\lbrace \text{diag}(\mathbf{PP}^H)\right\rbrace^{-\frac{1}{2}} \mathbf{\tilde{H}}^H\mathbf{\Sigma}^{-1}& \\\xrightarrow[K\to\infty, M\to\infty]{} \; \; &\frac{\sqrt{2}(M-K)^2}{\sigma_s
\sqrt{\sum_{i=1}^{K} \frac{1}{\sigma^2_i}}} \mathbf{I}_K,
\end{split}
\end{equation}
so that
\begin{equation}
\mathbf{G} \xrightarrow[K\to\infty,M\to\infty]{} \; \; \frac{2(M-K)}{\sigma_s\sqrt{\pi\sum_{i=1}^{K} \frac{1}{\sigma^2_i}}}
\mathbf{I}_K.
\end{equation}

The cross-covariance matrix of $\tilde{\mathbf{s}}$ and $\mathbf{s}$ becomes
\begin{equation}\label{RssDifVar}
\begin{split}
\mathbf{R_{\tilde{s}s}} \simeq &\sigma^2_s\frac{\sqrt{\rho_0}\mathbf{H}\mathbf{\tilde{H}}^H}{\sqrt{M\pi\sum_{i=1}^{K} \frac{1}{\sigma^2_i}}}\mathbf{\Sigma}^{-1} \\
\\ \xrightarrow[K\to\infty,M\to\infty]{} & \frac{2\sigma_s\sqrt{\rho_0}}{\sqrt{\pi}} \left(\frac{M}{K}-1 \right) \sqrt{\frac{K}{M}}
\frac{1}{\sqrt{\frac{1}{K} \sum_{i=1}^{K} \frac{1}{\sigma^2_i} }}
\mathbf{I}_K \\
= & \beta \mathbf{I}_K \; ,
\end{split}
\end{equation}
which simplifies to
\begin{equation}\label{eq_r_stildes}
\mathbf{R_{\tilde{s}s}}
 \xrightarrow[K\to\infty,M\to\infty]{} \frac{2\sqrt{\rho_0}}{\sqrt{\pi}} (M/K-1)\sqrt{K/M} \sigma\sigma_s\mathbf{I}_K
\end{equation}
for the case of path losses all equal to $\sigma$.
We observe that for large $M$ and $K$, $\tilde{s}_k$ and $s_l$ are uncorrelated for $k\neq l$, and hence
the multi-user interference disappears. In addition, we see that the signal of interest is received
with a gain of $\frac{\sqrt{2}}{\sigma_s}\beta$ which grows as $\sqrt{\frac{M}{K}}$.  Hence, the larger the value of $M/K$, the deeper the
received signal will be pushed into the correct decision region, and hence the lower the probability of
a decoding error in the presence of noise at the receiver.

\begin{figure}
	\centering
	\includegraphics[width=0.5\textwidth]{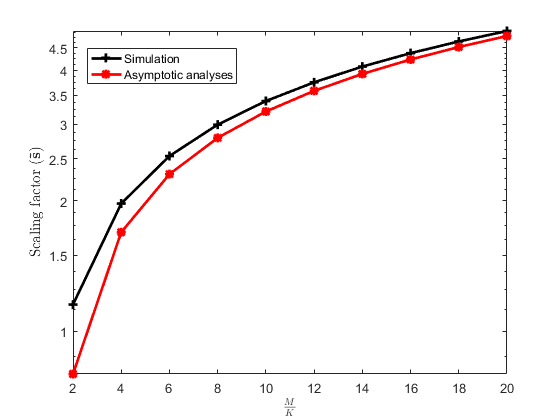}
	\caption{Asymptotic and simulated average scaling factor, $\frac{\sqrt{2}}{\sigma_s}\beta$ with respect to
$M/K$.}
	\label{fig_scale}
\end{figure}

In Fig.~\ref{fig_scale}, we plot the scaling factor, $\frac{\sqrt{2}}{\sigma_s}\beta$ found by simulation when averaging over all $K=20$ users and over $10^4$ channel realizations
for a case with $\sigma_i=\sigma=1,\forall i$. The simulation curve is
compared with the asymptotic formula in~(\ref{eq_r_stildes}), and shows
very good accuracy for $M/K > 10$.

\subsection{Asymptotic SQINR and Probability of Error}

As described in Section \ref{AsymptSigl}, the ZF precoder $\mathbf{PP}^H$ asymptotically becomes a diagonal matrix with $K$ non-zero diagonal elements which correspond to eigenvectors lying on the subspace spanned by $\mathbf{H}$. Using (\ref{eq_pph_asymptotic}) and the Law of Large Numbers \cite{Gubner2006Probability}, we have
\begin{equation}
\begin{split}
\mathbf{R_{dd}}=\mathbf{HR_{qq}H}^H = &\mathbf{\Sigma \tilde{H}R_{qq}\tilde{H}}^H\mathbf{\Sigma}\\ \xrightarrow[K\to\infty,M\to\infty]{} \; \; & 2\left( 1-\frac{2}{\pi}\right)(M-K)\mathbf{\Sigma}^2,
\end{split}
\end{equation}
The SQINR then becomes
\begin{equation}
SQINR_k \; \xrightarrow[K\to\infty,M\to\infty]{} \; \; \frac{\rho_0\frac{4(M-K)^2}{M\pi\sum_{i=1}^{K}\frac{1}{\sigma^2_i}}}{\rho_0\frac{2}{M}\left(1-\frac{2}{\pi}\right)(M-K)\sigma^2_k+\sigma^2_n},
\end{equation}
As before, assuming equally likely Gray-mapped QPSK symbols, we have the probability of error for the $k^{th}$ user as
\begin{equation}
\begin{split}
P_e=& Pr(\mathbb{Q}(r_k)\neq s_k) \simeq 2Q(\sqrt{SQINR_k}) \\ \xrightarrow[K\to\infty,M\to\infty]{} \; \; & 2Q\left(\sqrt{\frac{\rho_0\frac{4(M-K)^2}{M\pi\sum_{i=1}^{K}\frac{1}{\sigma^2_i}}}{\rho_0\frac{2}{M}\left(1-\frac{2}{\pi}\right)(M-K)\sigma^2_k+\sigma^2_n}}\right).
\end{split}
\end{equation}
In the case of equal channel gains,
\begin{equation}
SQINR_k \; \xrightarrow[K\to\infty,M\to\infty]{} \;\; \frac{\rho_0\frac{4\sigma^2(M-K)^2}{MK\pi}}{\rho_0\frac{2\sigma^2}{M}\left(1-\frac{2}{\pi}\right)(M-K)+\sigma^2_n},
\end{equation}
so that
\begin{equation}\label{error_pr_eqvar}
P_e  \simeq  2Q\left(\sqrt{\frac{\rho_0\frac{4\sigma^2(M-K)^2}{MK\pi}}{\rho_0\frac{2\sigma^2}{M}\left(1-\frac{2}{\pi}\right)(M-K)+\sigma^2_n}}\right).
\end{equation}

For high SNR scenarios, the $SQINR_k$ can be approximated with the signal to quantization and interference ratio ($SQIR_k$)
\begin{equation}
SQIR_k \simeq \frac{\frac{2}{\pi}(\frac{M}{K}-1)}{\left(1-\frac{2}{\pi}\right)
\left( \frac{1}{K} \sigma^2_k \sum_{i=1}^{K} \frac{1}{\sigma^2_i} \right)},
\end{equation}
and thus the probability of error will experience the following error floor
\begin{equation}
\begin{split}
P_e \simeq & 2Q(\sqrt{SQIR_k})\\\simeq & 2Q\left(\sqrt{\frac{\frac{2}{\pi}}{\left( 1-\frac{2}{\pi}\right) }}\sqrt{\frac{\frac{M}{K}-1}{\frac{\sigma^2_k}{K} \sum_{i=1}^{K}\frac{1}{\sigma^2_i}}}\right).
\end{split}
\label{eq_pe_asymptotic}
\end{equation}

\begin{figure}
	\centering
	\includegraphics[width=0.5\textwidth]{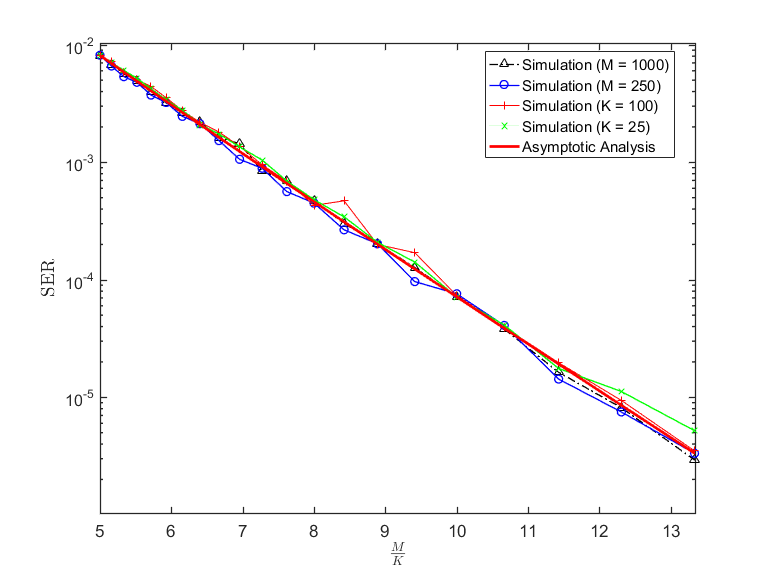}
	\caption{Variation of one-bit ZF precoding SER with the ratio $\frac{M}{K}$ in the noiseless case.}
	\label{fig_K_M}
\end{figure}

When the channel gains are equal, the high-SNR $SQIR_k$ is given by
\begin{equation}
SQIR_k \simeq \frac{\frac{2}{\pi}}{1-\frac{2}{\pi}}\left( \frac{M}{K}-1\right),
\end{equation}
so that the error floor becomes
\begin{equation}
P_e \simeq 2Q\left(\sqrt{\frac{\frac{2}{\pi}}{1-\frac{2}{\pi}}\left( \frac{M}{K}-1\right)}\right).
\label{eq_pe_equal_asymptotic}
\end{equation}
In all cases we note the critical dependence of the SQINR and probability of error on the quantity $M/K$;
in particular, the SQINR increases approximately linearly with $M/K$.
In Fig.~\ref{fig_K_M}, we have plotted the symbol error rate (SER) for the case of no additive noise
as a function of $M/K$ for various choices of $M$ and $K$ averaged over $10^6$ channel realizations. We see that the simulations match the analysis
very well, and illustrate the importance of the ratio $M/K$ on performance. Massive MIMO systems are
typically envisioned to operate with loading factors on the order of $M/K \simeq 10$, and we see that
for this case the SER is below $10^{-4}$, which bodes well for the use of the
quantized ZF precoder in practical scenarios.

\subsection{Simulations}

\begin{figure}
	\centering
	\includegraphics[width=0.5\textwidth]{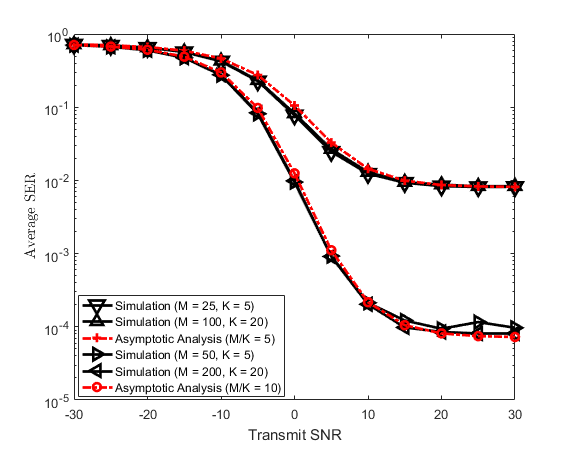}
	\caption{Variation of SER with transmit SNR, for varying number of users, $K$ and BS antennas, $M$.}
	\label{fig_k20}
\end{figure}
In Fig.~\ref{fig_k20}, we have plotted both the predicted and simulated SER for one-bit quantized ZF precoding at the BS for varied number of users
as a function of the transmit SNR, $\rho_0$, as described in Section~\ref{OneBitPrecodeModel}. Again, $10^6$ channel realizations were used to generate the results, assuming equal channel gains, $\sigma_i=1,\forall i=1, \ldots, K.$ We note the excellent match between the simulations and analytical approximation in ~(\ref{error_pr_eqvar}), which validates our analysis. As expected, we observe that the SER approaches an error floor at high SNR; for example, with $M\sim 10K$, the SER floor is of the order of $10^{-4}$. To see how the analysis holds for non-asymptotic values of $K$ and $M$, we have also performed the simulation for the case of $K=5$ users. We observe that our asymptotic analysis is accurate even for this non-asymptotic value. The results of simulation are similar for $K=5$ and $K=20$ for both $M/K = 5,10$.  This result reinforces the observation that performance is governed by the ratio $M/K$, independent of their specific values.

\begin{figure}
	\centering
	\includegraphics[width=0.5\textwidth]{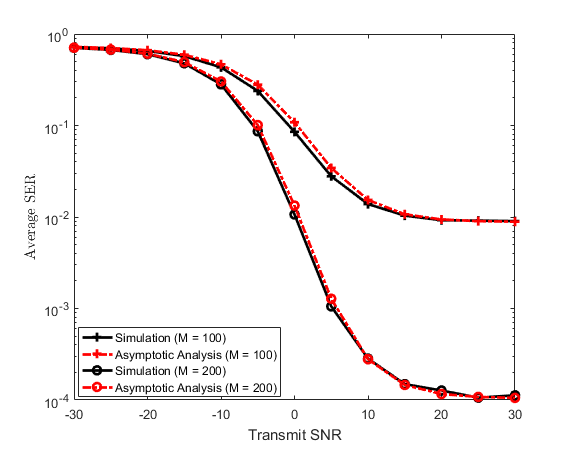}
	\caption{Variation of SER with transmit SNR for $K=20$ users with unequal channel gains and varying values of BS antennas $M$.}
	\label{fig_var_diff}
\end{figure}
In Fig~\ref{fig_var_diff}, we plot the average SER for the quantized ZF precoding scenario, again with $K=20$. Unlike the previous example, we have assumed here that
the users have unequal channel gains; in this simulation, the square of the gains were chosen as independent log-normal random variables, such that $\ln(\sigma^2_i)\sim \mathcal{N}(\mu, \sigma^2), \, \forall \, i=1, \ldots, K$ with parameters $\sigma=0.125$ and $\mu=-\sigma^2/2$, so that for large enough $K$, $\frac{1}{K} \sum_{k=1}^K \sigma_k^2
\rightarrow e^{\mu+\frac{\sigma^2}{2}}=1$. $10^5$ channel realizations were used to generate the results in this case. For this case also, we can see that the SER approaches an error floor, of the order of $10^{-4}$ when $M \sim 10K$ for high SNR. The simulations agree very well with the asymptotic analysis.

\begin{figure}
	\centering
	\includegraphics[width=0.5\textwidth]{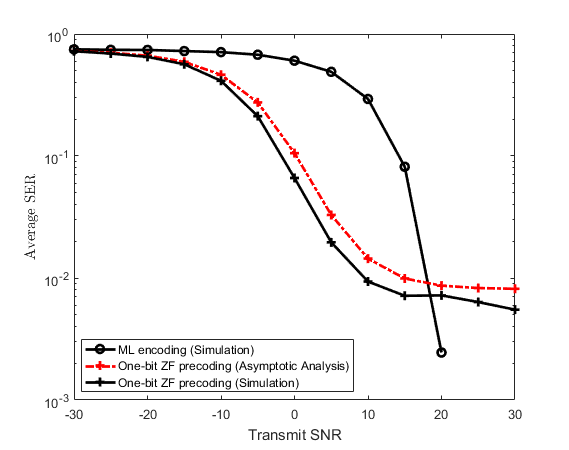}
	\caption{Variation of SER versus transmit SNR for $K=2$ users and $M=10$ BS antennas for one-bit ZF precoding and ML encoding for equal channel gains.}
	\label{fig_ZF_ML}
\end{figure}

In Fig.~\ref{fig_ZF_ML} we compare the ML encoding approach (\ref{eq_ml}) with the quantized ZF precoder.  Due to the
complexity of the ML encoder, we perform the simulation for the relatively small values $M=10$ and $K=2$. The gains of the channel rows are assumed to be equal: $\sigma_1=\sigma_2=1$, and $10^4$ channel realizations were generated to compute the results. While the ML encoder is superior at high SNR, there is a broad range of low- to medium-SNRs where the simple quantized ZF precoder
provides significantly better performance. The low- to medium-SNR range is of particular interest for massive MIMO
implementations, and thus the quantized ZF precoder is an attractive simple approach for such scenarios.

%
%
%

\begin{figure}
	\centering
	\includegraphics[width=0.5\textwidth]{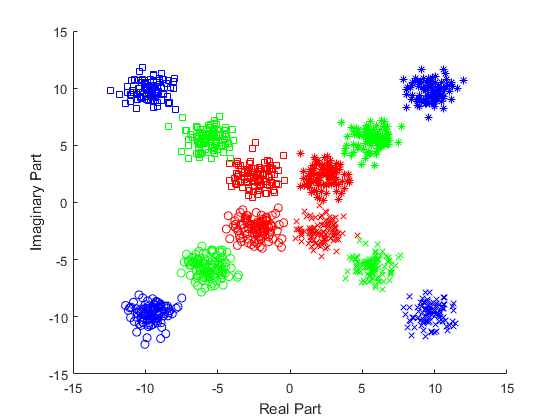}
	\caption{Plot of the received vector in a noiseless scenario, for $K=4$, and different values of $M$, $M=20\text{ (red)}, 100\text{ (green)}, 300\text{ (blue)}$. Different components of the received vector are shown as different symbols.}
	\label{figure_Scatterplot}
\end{figure}

For the next example, we take the special case of $K=4$ and assume that each user is being sent a different symbol. In particular, the desired user symbols are chosen to be $\mathbf{s}=[1+j, 1-j, -1+j, -1-j]^T$. In Fig.~\ref{figure_Scatterplot}, we plot the signals received by each of the four users (different symbol for each user) and for $M=20, 100, 300$ (different color for each $M$), assuming no receiver noise (the only source of error here is the quantization at the transmitter). The simulations have been performed over $10^2$ independent channel realizations.  We clearly see that as the value of $M$ increases, the average distance of the symbols from the decision boundary increases, which indicates an increased robustness with respect to additive noise with increasing $M/K$.

\section{Bussgang Adapted One-Bit ZF Precoder}

In this section, we use insights gained from our analysis of the quantized ZF precoder to improve its performance.
We have seen that the cross-covariance matrix of $\tilde{\mathbf{s}}$ and intended vector $\mathbf{s}$ (\ref{RssDifVar}) reduces asymptotically with $M$ and $K$ to a
diagonal matrix when using the ZF precoder.
Our objective in the approach we present here is to enforce this diagonal feature of
$\mathbf{H}\left\{ {\rm diag}{(\mathbf{PP}^H})\right\}^{-1/2}\mathbf{P}$ for any value
of $M$ and $K$, by an improved choice of $\mathbf{P}$.

\subsection{Principle}

For $\mathbf{H}\left\{ {\rm diag}{(\mathbf{PP}^H})\right\}^{-1/2}\mathbf{P}$ to be diagonal, we must choose $\mathbf{P}$ so that
\begin{equation}
\left\{ {\rm diag}{(\mathbf{PP}^H})\right\}^{-1/2}\mathbf{P}
= \mathbf{H}^\dagger \mathbf{D},
\label{eq_diag}
\end{equation}
where $\mathbf{H}^\dagger=\mathbf{H}^H(\mathbf{HH}^H)^{-1}$ and $\mathbf{D}$ represents some real-valued positive diagonal matrix.
For this to be true, we can see that $\mathbf{P}$ must satisfy,
\begin{equation}
\mathbf{P}=\mathbf{\Lambda} \mathbf{H}^\dagger \mathbf{D},
\label{eq_pdiag}
\end{equation}
with $\mathbf{\Lambda}=\left\{ {\rm diag}{(\mathbf{PP}^H})\right\}^{1/2}$.
Using (\ref{eq_pdiag}) in (\ref{eq_diag}), the following condition must then hold:
\begin{equation}
\mathbf{\Lambda}^{-1/2} \{ \text{diag}(\mathbf{H}^\dagger \mathbf{D}^2 (\mathbf{H}^\dagger)^H)\}^{-1/2} \mathbf{\Lambda}^{1/2}\mathbf{H}^\dagger=\mathbf{H}^\dagger,
\nonumber
\end{equation}
which simplifies to
\begin{equation}
\text{diag}(\mathbf{H}^\dagger \mathbf{D}^2 (\mathbf{H}^\dagger)^H) = \mathbf{I}_M \; .
\label{eq_d2}
\end{equation}
Note that $\mathbf{\Lambda}$ has vanished from (\ref{eq_d2}) indicating that it does not affect the signal of interest, and thus
we take $\mathbf{\Lambda} = \mathbf{I}_M$ for simplicity.

Denoting $\mathbf{T}=\mathbf{H}^\dagger$, (\ref{eq_d2}) becomes $\text{diag}(\mathbf{TD}^2\mathbf{T}^H)=\mathbf{I}_M$.
It can be written with respect to the diagonal entries of $\mathbf{D}=\text{Diag}(d_1, d_2, \ldots, d_K)$ as
\begin{equation}
\begin{bmatrix}
|t_{11}|^2 \ldots  |t_{1K}|^2\\
|t_{21}|^2  \ldots |t_{2K}|^2 \\
\ldots\\
|t_{M1}|^2 \ldots |t_{MK}|^2
\end{bmatrix}
\begin{bmatrix}
d^2_{1}\\
d^2_{2}\\
\hdotsfor{1} \\
d^2_{K}
\end{bmatrix}
=\mathbf{1}_M \; ,
\end{equation}
where $\mathbf{1}_M$ is an $M \times 1$ vector of ones.
Now define the $M\times K$ matrix
\begin{equation}
\mathbf{\tilde{T}}=\begin{bmatrix}
|t_{11}|^2 \ldots  |t_{1K}|^2\\
|t_{21}|^2  \ldots |t_{2K}|^2 \\
\ldots\\
|t_{M1}|^2 \ldots |t_{MK}|^2
\end{bmatrix}
\nonumber
\end{equation}
so that
\begin{equation}
\begin{bmatrix}
d^2_{1}\\
d^2_{2}\\
\hdotsfor{1} \\
d^2_{K}
\end{bmatrix}=(\mathbf{\tilde{T}}^H\mathbf{\tilde{T}})^{-1}\mathbf{\tilde{T}}^H\mathbf{1}_M.
\label{eq_d2found}
\end{equation}
Solving (\ref{eq_d2found}), we can immediately deduce $\mathbf{D}$.

For the precoder we obtain solving (\ref{eq_diag}), the cross-covariance matrix between $\mathbf{\tilde{s}}$ and $\mathbf{s}$ is
\begin{equation}
\mathbf{R_{\tilde{s}s}}=\frac{\sqrt{\rho_0}\mathbf{HFP}}{\sqrt{M}}\mathbf{R_{ss}}=
\sqrt{\frac{2\rho_0}{M\pi}}\sigma_s\mathbf{D} \; ,
\end{equation}
indicating that all multi-user interference has been canceled.
The received signal can thus be described as
\begin{equation}
\mathbf{r}=\sqrt{\frac{2\rho_0}{M\pi}}\frac{1}{\sigma_s}\mathbf{Ds}+\sqrt{\frac{\rho_0}{M}}\mathbf{Hq}+\mathbf{n}.
\end{equation}
The $SQINR$ for the $k^{th}$ received symbol will be
\begin{equation}
SQINR_k=\frac{\frac{2\rho_0}{M\pi}d^2_k}{\frac{\rho_0}{M}[\mathbf{H}\mathbf{R_{qq}}\mathbf{H}^H]_{kk}+\sigma^2_n} \; ,
\end{equation}
and the SER for the $k^{th}$ user for QPSK
modulation and equally likely signaling is
\begin{equation}
P_e\simeq 2Q\left(\sqrt{\frac{\frac{2\rho_0}{M\pi}d^2_k}{\frac{\rho_0}{M}[\mathbf{H}\mathbf{R_{qq}}\mathbf{H}^H]_{kk}+\sigma^2_n}}\right) 
\end{equation}

\subsection{Proposed Algorithm}

\begin{figure}
	\centering
	\includegraphics[width=0.5\textwidth]{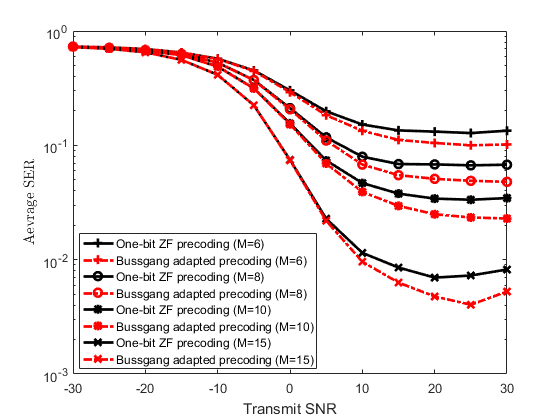}
	\caption{Variation of SER for one-bit quantized Bussgang adapted and
ZF precoding versus transmit SNR for $K=3$ users and varying values of $M$.}
	\label{fig_bussgang}
\end{figure}

\begin{figure}
	\centering
	\includegraphics[width=0.5\textwidth]{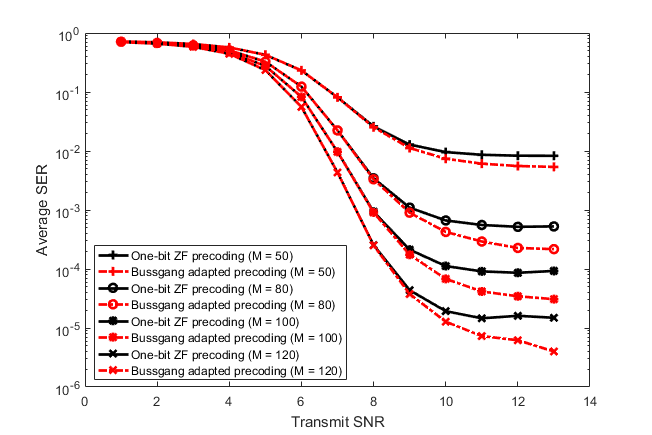}
	\caption{Variation of SER for one-bit quantized Bussgang adapted and
		ZF precoding versus transmit SNR for $K=10$ users and varying values of $M$.}
	\label{fig_bussgangK_10}
\end{figure}

Since we know $\mathbf{H}$, it is easy to check whether or not the Bussgang-adapted precoded
vector induces fewer errors at the receiver than the ZF precoder.
Therefore, we propose the algorithm shown in the table labeled ``Algorithm 1'' below. For a given vector $\mathbf{s}$ and channel matrix, $\mathbf{H}$, the initial ZF precoding matrix $\mathbf{P}=\mathbf{H}^\dagger$ is estimated. It is then checked whether the channel output of the precoded matrix, when subjected to quantization, gives an output that is equal to the vector $\mathbf{s}$. If it is, then the precoding matrix is taken to be $\mathbf{P}=\mathbf{H}^\dagger$. However, if it is not, we take an auxiliary precoding matrix, $\mathbf{\tilde{P}}=\mathbf{H}^\dagger\mathbf{D}$, where $\mathbf{D}$ is calculated from (\ref{eq_d2found}), and check which of the precoding matrices, out of the ZF precoder and the auxiliary precoding matrix, generate a lower number of errors, when the precoded vector is sent across the channel in a noiseless condition and decoded at the receiver. The precoding matrix that offers the lowest number of errors is selected for precoding.
In Fig.~\ref{fig_bussgang} and Fig.~\ref{fig_bussgangK_10}, we compare the SER performance of the Bussgang adapted precoding
approach with quantized ZF precoding as a function of the transmit SNR for a case with $K=3$ and $K=10$ users respectively. We
see that the new algorithm achieves a lower error floor in all cases compared with ZF precoding. Also, we can observe that for a fixed number of users, $K$, the improvement with respect to the ZF precoding increases slightly with increasing number of BS antennas, $M$.

\begin{algorithm}
	\textbf{Input}: $\mathbf{s}, \mathbf{H}$\\
		Let $\mathbf{P}=\mathbf{H}^\dagger$\\
	\If{$\mathbb{Q}(\mathbf{H}\mathbb{Q}(\mathbf{Ps}))\neq \mathbf{s}$}{
		Let $\mathbf{\tilde{P}}=\mathbf{H}^\dagger\mathbf{D}$, where $\mathbf{D}$ can be found using equation (\ref{eq_d2found})\\ \If{$\|\mathbf{1}_{\mathbb{Q}(\mathbf{H}\mathbb{Q}(\mathbf{Ps}))\neq \mathbf{s}}\|_1>\|\mathbf{1}_{\mathbb{Q}(\mathbf{H}\mathbb{Q}(\mathbf{\tilde{P}s}))\neq \mathbf{s}}\|_1$}{
			$\mathbf{P}=\mathbf{\tilde{P}}$
			}
					}
\textbf{Output}: Resulting precoded vector is $\mathbf{x}_P=\mathbf{Ps}$
\caption{Bussgang adapted precoding algorithm}
\end{algorithm}

\section{Conclusion}

We have studied the use of quantized linear precoding for the massive MIMO downlink with one-bit DACs. We derived closed form expressions for the SQINR and SER for any linear precoder using the Bussgang decomposition. We provided an analysis to show that asymptotically in the number of antennas $M$ and the number of users $K$, the algorithm yields signals at the user terminals that are scaled versions of the desired symbols, with the scaling increasing proportionally to $\sqrt{M/K}$. Simulations show that the algorithm outperforms the ML encoder for low to moderate SNRs for the scenario considered. We also presented a modified version of the quantized ZF precoder that yields lower SERs at high SNR.

\section*{Acknowledgment}
The research was supported by the National Science Foundation under Grant ECCS-1547155, and by the Institute for Advanced Study at the Technische Universit\"at M\"unchen, funded by the German Excellence Initiative and the European Union Seventh Framework Programme under grant agreement No. 291763, and by the European Union under the Marie Curie COFUND Program, and by the French CNRS.


\bibliographystyle{ieeetr}
\bibliography{journbib}

\end{document}